\documentclass[review]{elsarticle}
\usepackage{float}
\usepackage{tabulary,booktabs}
\usepackage{hyperref}
\usepackage{multirow}
\usepackage[utf8]{inputenc}
\usepackage[lined,ruled,commentsnumbered]{algorithm2e}
\usepackage{caption} 
\usepackage{commath}
\usepackage{amssymb} 
\usepackage{subcaption}

\journal{Journal of Information Processing and Management}

\begin{document}

\begin{frontmatter}

\title{Fair Multi-Stakeholder News Recommender System with Hypergraph ranking}

\author[mymainaddress,mysecondaryaddress]{Alireza Gharahighehi}
\author[mymainaddress,mysecondaryaddress]{Celine Vens}
\author[mymainaddress,mysecondaryaddress]{Konstantinos Pliakos}

\address[mymainaddress]{Itec, imec research group at KU Leuven, Kortrijk, Belgium}
\address[mysecondaryaddress]{KU Leuven, Campus KULAK, Dept. of Public Health and Primary Care, Kortrijk, Belgium}

\begin{abstract}
Recommender systems are typically designed to fulfill end user needs. However, in some domains the users are not the only stakeholders in the system. For instance, in a news aggregator website users, authors, magazines as well as the platform itself are potential stakeholders. Most of the collaborative filtering recommender systems suffer from popularity bias. Therefore, if the recommender system only considers users' preferences, presumably it over-represents popular providers and under-represents less popular providers. To address this issue one should consider other stakeholders in the generated ranked lists. In this paper we demonstrate that hypergraph learning has the \textit{natural} capability of handling a multi-stakeholder recommendation task. A hypergraph can model high order relations between different types of objects and therefore is naturally inclined to generate recommendation lists considering multiple stakeholders. We form the recommendations in time-wise rounds and learn to adapt the weights of stakeholders to increase the coverage of low-covered stakeholders over time. The results show that the proposed approach counters popularity bias and produces fairer recommendations with respect to authors in two news datasets, at a low cost in precision. 

\end{abstract}


\end{frontmatter}

\section{Introduction}

Recommender systems (RSs) are ubiquitous in digital services and rank items based on user preferences. In many applications there is an abundant number of unique items but each user is only interested in a small portion of them. The task of an RS is to assist users in finding the items they would likely be interested in. This is carried out by filtering and ranking vast item collections, typically taking into account user past preferences as well as item content information. A specific case is news aggregator websites, where numerous news articles are published every day and an RS should filter and rank the most relevant articles for users. In RSs where users are the only stakeholders, articles are ranked based on similarity (content-based filtering), based on user interactions (collaborative filtering), or a combination of both (hybrid approach). However, in news websites there are also other stakeholders such as authors, magazines and the platform itself that should be considered in recommendation. 

Ignoring other stakeholders' preferences causes unfair and biased recommendations that have a negative impact on the ignored parties of the system. For instance, in news recommendations, if the RS only focuses on the user preferences it is highly likely that it reinforces the popularity bias~\cite{park2008long} and consequently ignores the articles of less popular authors/journalists. Recommending only very popular articles also bolsters the filter bubble phenomenon. This may lead to limited (unfair) exposure of less popular journalists even if their articles are relevant for a group of readers. This unfair exposure may have negative consequences for these stakeholders and they may gradually lose their trust in the system which will impact other parties and the whole system as well~\cite{boratto2020interplay}. Most collaborative filtering-based RSs serve the users who like popular items very well, but they can not fulfill the preferences of the users who are interested in less-popular and niche themes~\cite{abdollahpouri2020addressing}. A multi-stakeholder news RS generates recommendation lists not only based on users' preferences but also based on other stakeholders in the system. To have the capability of considering multiple stakeholders, an RS needs to model the relations between different stakeholders and objects in the system. For instance, in news aggregator websites a multi-stakeholder RS should consider relations between users, magazines, authors, articles, tags, etc.

In this paper, we propose to use hypergraph learning to include multiple stakeholders in recommendations. We build a unified hypergraph to represent multiple types of objects and their complex relations in the context of news recommendation. We show that hypergraph learning has the propensity of addressing multi-stakeholder news recommendation. A hypergraph is a generalization of a graph where instead of edges between pairs of objects, multiple types of objects, which are represented as vertices (nodes), are linked via hyperedges. This way, the RS benefits from capturing high order relations between the objects. This setting naturally fits in a multi-stakeholder RS, as different stakeholders can be represented as different types of vertices in the hypergraph. Therefore, the relatedness between different stakeholders (e.g. between a specific user and several authors in the dataset) can be directly measured. The impact of each vertex type (i.e. stakeholder) in the creation of a recommendation list can be also directly adjusted.    

This study is an extension of our previous work~\cite{hyper_news_multi} where we briefly discussed the employment of hypergraph ranking as a multi-stakeholder RS. We also demonstrated a naive solution where we improved the coverage of a specific stakeholder by manually introducing corresponding weights in the construction of the query vector (i.e. a vector where every cell corresponds to an object, such as a user, article, author, etc.). Different from \cite{hyper_news_multi}, here we propose a temporal-aware learning approach where the weights of the different stakeholders get dynamically updated to increase recommendation coverage and fairness without jeopardizing precision. More specifically, we perform the recommendation task in simulated rounds, making it applicable to the news domain by taking into account the temporal factor. Furthermore, we propose an approach, where in each round, the weights of under-represented stakeholders get dynamically adapted in the query vector, taking into account both coverage and relevance. This way, we constantly increase the coverage of under-represented stakeholders while preventing a significant drop in precision. This addresses the low and unfair visibility of under-represented stakeholders in the recommendation lists and mitigates popularity biased recommendations. To the best of our knowledge, this is the first approach that addresses fairness in a multi-stakeholder RS setting using hypergraph learning. Similar query adaptation approaches can be used to address other issues in RSs, such as the filter bubble phenomenon and cold start problem. We briefly illustrate how the filter bubble phenomenon can be addressed by diversifying the recommendations and leave the cold start problem for future work.

In the following, related studies about multi-stakeholder RSs, fairness and diversification in recommendation, as well as hypergraph learning are presented in Section~\ref{sec2}. Next, in Section~\ref{sec3}, the proposed approach to mitigate biased and unfair recommendations from a multi-stakeholder perspective is explained. In Section~\ref{sec4}, two news datasets are described and the experimental settings in designing and testing the proposed model are discussed. Next, we compare the proposed method with other approaches from the literature. The obtained results are presented and thoroughly discussed in Section~\ref{sec5}. Finally, we draw conclusive remarks in Section~\ref{sec6}.    

\section{Related work}
\label{sec2}
The related studies can be categorized in four categories: multi-stakeholder RSs, fairness in multi-stakeholder recommendations, diversity in recommendations, and hypergraph-based RSs. 

\subsection{Multi-stakeholder Recommender Systems}
In many applications, including news aggregators, users are not the only stakeholders and there are some other parties that have stake in recommendations. A multi-stakeholder RS is a system that ranks items not only based on the end user’s preference, but also on the interests of other stakeholders~\cite{abdollahpouri2019beyond}. 

Studies that consider multiple stakeholders in recommendations generally follow one of the following two methodological approaches: considering the preferences of multiple stakeholders by including them in the core optimization phase (model-based) or through a post-processing step (re-ranking)~\cite{abdollahpouri2019beyond}. In the model-based approach multiple stakeholders can be included by optimizing over multiple objective functions~\cite{zheng2019preference}, by adding regularization terms to the objective function~\cite{boratto2020interplay,abdollahpouri2019incorporating,burke2018balanced}, or by adding constraints to include multiple criteria in the optimization~\cite{surer2018multistakeholder,svore2011learning,jambor2010optimizing}. The post-processing approach relies on re-ranking the initial recommendation list of a base RS. Burke \textit{et al.}~\cite{burke2016towards} proposed a re-ranking approach to increase the number of satisfied stakeholders in recommendations. Their coverage-oriented algorithm maximizes the number of included providers while still recommending relevant items to the end-users. 

\subsection{Fairness in multi-stakeholder recommendations}
Fairness in recommendation is a subjective matter and its definition depends on a pre-determined fairness objective. It has been mainly studied as the parity and equity of recommendations for individuals or groups of users~\cite{dwork2012fairness}. From the multi-stakeholder perspective, fairness has received far less attention in the literature. In the study by Burke \textit{et al.}~\cite{burke2016towards}, having more satisfied stakeholders in the system brings more utility for the whole platform. From the providers' perspective, satisfaction relies on being covered in recommendations in the course of time. Abdollahpouri \textit{et al.}~\cite{abdollahpouri2020addressing} investigated whether RSs transfer and reinforce the popularity bias of interactions in their recommendations. This bias leads to over-representation of items with more interactions over less popular items~\cite{abdollahpouri2019popularity}. They proposed a re-ranking approach to resolve this bias by having a balance between accuracy and fairness using the relevance scores and the popularity of the items in recommendations. They also assessed the effect of this adjustment from the providers' perspective. However, they did not explicitly address the popularity bias on the providers' side in the model. Boratto \textit{et al.}~\cite{boratto2020interplay} mitigated the bias towards over-represented providers in two steps. First, they preprocessed the training set by up-sampling the interactions in the training data to reduce the low converge of under-represented providers. Second, they added a regularization term to the optimization task to have a better balance between relevance and fairness. 


\subsection{Diversity in recommendations}
The concept of diversity was first introduced in information retrieval to disambiguate users' queries~\cite{kaminskas2017diversity}. An example of this ambiguity is the search term \textit{jaguar} which can be the animal, the brand of car or a type of guitar. A diversified retrieved list of documents is more likely to cover the real intent of the query. In RSs diversity is meant to overcome the filter bubble phenomenon. For instance, in a news aggregator website, a recommended list that contains news about politics, entertainment and sports, has higher diversity than a recommended list containing only political news. There are two main approaches to enhance the diversity of recommendations: diversity modeling~\cite{su2013set,wasilewski2016incorporating,shi2012adaptive} and re-ranking approaches~\cite{vargas2011intent,vargas2014improving,barraza2015xplodiv}. The re-ranking studies mostly follow the greedy approach using maximal marginal relevance~\cite{carbonell1998use} to have a better trade-off between different objective functions such as accuracy and diversity.

\subsection{Hypergraph-based RSs}
Hypergraph ranking has been used for recommendation tasks in various application fields. Bu \textit{et al.}~\cite{bu2010music} utilized hypergraph learning to perform music recommendation. They specifically built a unified hypergraph model that captured the relations between users, songs, albums and musicians, effectively modeling the high order relations between them. Pliakos \textit{et al.}~\cite{pliakos2014simultaneous} also used a unified hypergraph model boosted by group sparsity optimization. They encapsulated the high order connections between users, images, tags, and geo-tags for tag recommendation. Moreover, Li \textit{et al.}~\cite{li2013news} built a news recommender system based on a hypergraph. They constructed a model based on users, news articles, topics and named entities. In \cite{Wang:2020}, hypergraphs were combined with multiple convolutional layers for next item recommendation in e-commerce. In another e-commerce RS application \cite{Mao:2019}, hypergraph ranking was employed modeling relations between users, items (e.g. restaurants), item-related attributes (e.g. dinner, alcohol), and item-related tips/tags. In all these methods the users were the one and only stakeholder of interest. Although hypergraph ranking models are very promising and have been used to model relations between different types of objects, to the best of our knowledge, they have never been studied from the perspective of multi-stakeholder recommendation.

\section{Methodology}
\label{sec3}
In this section we thoroughly describe the proposed methodology. In Section~\ref{sec:h_theory}, we present the theoretical background on hypergraph ranking. In Section~\ref{sec:query}, we elucidate our proposed approach to dynamically handle coverage and fairness through query vector adaptation. Finally in Section~\ref{sec:diveristy} we show how the query vector can be adapted to bring more diversified recommendations.

\subsection{Hypergraph ranking}
\label{sec:h_theory}
A hypergraph $G(V,E,w)$ consists of a set of vertices $V = \{v_1,v_2\cdots,v_{|V|}\}$ and hyperedges $E= \{e_1,e_2\cdots,e_{|E|}\}$. The weight function $w: E \rightarrow \mathbb{R}$ is assigned to $E$ differentiating the impact of each hyperedge $e_i \in E$. A hyperedge can model connections between multiple vertices of the hypergraph. For our news setting the set of vertices $V$ is composed of users, news articles, authors and topics. Let $\mathbf{H}$ of size $|V|\times|E|$ be the incidence matrix of the hypergraph, where $H{(v,e)=1}$, if node $v$ is in hyperedge $e$ and $0$ otherwise. Typically, the matrix $\mathbf{H}$ is very sparse and large. The vertex and hyperedge degrees are defined based on $\mathbf{H}$ as follows:  
\begin{eqnarray}
\label{eq:dmatdef}
& & \;\; \delta(v)=\sum_{e \in E}w(e)H(v,e) \\
    \nonumber \\
& & \;\; \delta(e)=\sum_{v \in V}H(v,e). 
\label{eq:dmatdef2}
\end{eqnarray}

Next, the diagonal matrices $\mathbf{D}_{v}$, $\mathbf{D}_{e}$, and $\mathbf{W}$ are defined. In more detail, $\mathbf{D}_{v}$ is the vertex degree matrix of size $|V|\times|V|$ and $\mathbf{D}_{e}$ is the hyperedge degree matrix of size $|E|\times|E|$. $\mathbf{W}$ is the $|E|\times|E|$ matrix that contains the hyperedge weights in its diagonal. Here, for simplicity we set $\mathbf{W} = \mathbf{I}$, where $\mathbf{I}$ is the identity matrix. Although there are approaches to adjust or optimize the hyperedge weights (e.g. \cite{pliakos2014weight}), for the sake of simplicity, here we assign equal weights to all the hyperedges. In our setting, we build our RS on an undirected and non-uniform hypergraph, where the latter means that each $\delta(e_i)$ can be different. The exact structure of our model is demonstrated in Section \ref{sec4_2}. 

Let the adjacency matrix $\mathbf{A} \in \mathbb{R}^{|V|\times|V|}$ be formed using Eq.\ref{eq1}:

\begin{equation}
\mathbf{A}=\mathbf{D}_{v}^{-1/2}\mathbf{H}\mathbf{W}\mathbf{D}_e^{-1}\mathbf{H}^{T}\mathbf{D}_{v}^{-1/2}.
\label{eq1}
\end{equation}

\noindent Matrix $\mathbf{A}$ is symmetric and each item $A_{ij}$ reflects the relatedness between vertices $i$ and $j$. A higher $A_{ij}$ value indicates a stronger relation between vertices $i$ and $j$. As it can be seen from (\ref{eq1}), $\mathbf{D}_{v}^{-1/2}$ and $\mathbf{D}_{e}^{-1}$ act as inverse cardinality weights for corresponding vertices and hyperedges, respectively. This contributes to the reduction of the effect of very popular objects, such as articles that appear in too many hyperedges. Next, the hypergraph Laplacian matrix is computed as $\mathbf{L}=\mathbf{I}-\mathbf{A}$ \cite{zhou2006learning}. 

The objective here is to find a ranking (score) vector $\mathbf{f}\in \mathbb{R}^{|V|}$ that minimizes the following cost function~\cite{bu2010music}:

\begin{equation}
Q(\mathbf{f})=\frac{1}{2}\mathbf{f}^{T} \mathbf{L} \mathbf{f}+\vartheta||\mathbf{f} - \mathbf{y}||_2^2.
\label{eq2}
\end{equation}

The first term in equation (\ref{eq2}) $\Omega(\mathbf{f}) = \mathbf{f}^{T}\mathbf{L}\mathbf{f}$ requires all vertices with very similar value in vector $\mathbf{f}$ to be strongly connected (i.e. share many hyperedges). The second term of the same equation is an $\ell_2$ regularization norm between the ranking vector $\mathbf{f}$ and the query vector $\mathbf{y} \in \mathbb{R}^{|V|}$. This drives the optimization problem to a solution where $\mathbf{f}$ can not be entirely different to $\mathbf{y}$. The solution of the optimization problem $argmin_\mathbf{f}Q(\mathbf{f})$, yields the following optimal score (ranking) vector $\mathbf{f}^{\ast}$: 

\begin{equation}
\mathbf{f}^{\ast} = \frac{\vartheta}{1+\vartheta}\bigl(\mathbf{I} - \frac{1}{1+\vartheta} \mathbf{A} \bigr)^{-1}\mathbf{y}.
\label{eq_fvector}
\end{equation}

\noindent Finally the scores of vertices in $\mathbf{{f}}^{\ast}$ that correspond to the articles are extracted as their relevance scores.  

\subsection{Query adaptation for coverage and fairness}
\label{sec:query}
To generate the recommendation list for user \(u\) in a regular recommendation task, one sets the corresponding value in the query vector to one ($y_{u}=1$) and all the other values to zero. Unlike using typical one-spot evaluation in RSs, we conduct simulated rounds to evaluate recommendations in a time-sensitive way~\cite{burke2016towards,burke2010evaluating}. In this regard, we set evaluation moments within fixed time intervals (e.g. every two weeks) and evaluate the model performance over time. We propose a method that learns to dynamically adapt the weights of stakeholders in the query vector based on the recommendations of the model in the previous rounds. This way we enhance the coverage of stakeholders, focusing on the under-represented providers in the previous rounds. 

In this study, we specifically enhance fairness toward authors in news recommendation tasks. The weight of each author in round \textit{t} is calculated based on Eq.~\ref{eq3}:

\begin{equation}
w^{t}_{i} = max(0,p_{i}^{t}-c^{t-1}_{i}) 
\label{eq3}
\end{equation}

\noindent where \(w^{t}_{i}\) is the weight of author \(i\) in round \(t\), \(p_{i}^{t}\) is the frequency of author \(i\) in the training set and \(c^{t-1}_{i}\) is the coverage ratio of author \(i\) in previous rounds till round \(t-1\). The author's frequency is the ratio of interactions in the training set that belong to articles written by author \(i\) and the author's coverage is the ratio of recommendations in previous rounds that contain the author's articles. 

These weights are used to avoid over-representation of popular authors and to promote under-represented authors, by calibrating their weights based on their popularity in the training set. However, focusing only on the coverage of under-represented authors may deteriorate the performance of recommendations in terms of accuracy. To address this issue we scale these weights based on the relevance of authors for a specific user. To calculate the relevance scores of authors for each user we first query the constructed hypergraph for that specific user ($y_{u}=1$) and then retrieve the authors' scores (\(\mathbf{{f}}^{\ast}(authors)\)) from the score vector $\mathbf{{f}}^{\ast}$ (Eq. \ref{eq_fvector}).

\begin{equation}
r^{t}_{j}=normalize(\hat f_{j}(authors)). 
\label{eq:relevance}
\end{equation}

\begin{figure}
\centering
  \includegraphics[width=\linewidth]{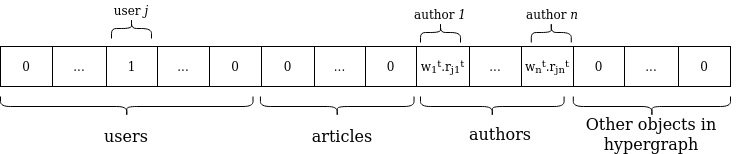}

\caption{The adapted query vector for user \textit{j} with authors' weights}
\label{fig:query}
\end{figure}

After normalizing these authors' relevance scores, we take the dot product of the authors' coverage weights (\(w^{t}_{i}\)) introduced in Eq.\ref{eq3} and their relevance scores (\(r^{t}_{i}\)) to calculate the final weights of the authors in the query vector. These calculated weights are used instead of zero values for the authors in the query vector (Figure~\ref{fig:query}). Algorithm~\ref{alg:simulation} illustrates the aforementioned steps. This approach can be easily extended to consider multiple stakeholders such as users, authors, publishers, named entities, etc.  

\begin{algorithm}
    \caption{Fair news recommendation using hypergraph}
    \SetKwData{Left}{left}\SetKwData{This}{this}\SetKwData{Up}{up}
    \SetKwFunction{Union}{Union}\SetKwFunction{FindCompress}{FindCompress}
    \SetKwInOut{Input}{input}\SetKwInOut{Output}{output}

    \Input{$d_{1},d_{2},...,d_{n}$ \tcp{ splitting the dataset to $n$ rounds}}
    \Output{top k recommended articles per round for each user}
    \BlankLine
    \emph{$n \gets len({d_{1},d_{2},...,d_{n}})$}\;
    \emph{$t\gets1$}\;
    \emph{$w^{t} , r^{t}\gets$ initialize with zero vectors}\;
    \While{$t\not=n$}{
        $tr_{t},ts_{t}\gets$ split $d_{t}$ to train and test sets\;
        $model\gets$ construct the hypergraph using $tr_{t}$ \;
        \If{$t\not=1$}{
            $authors\gets$ set of authors in $tr_{t}$\;
            \For{$i\in authors$}{$w^{t}_{i} \gets max(0,p_{i}^{t}-c^{t-1}_{i})$\tcp{ Eq.\ref{eq3}}}
            $users\gets$ set of users in $tr_{t}$\;
            \For{$j\in users$}{$r^{t}_{j} \gets $ query $model$ for authors' relevance scores \tcp{ Eq.\ref{eq_fvector}\&\ref{eq:relevance}}}}
   recommendations $\gets$ recommend news based on $model$, $w^{t}$ and $r^{t}$\;
   $topk^{t}\gets$ return $topk^{t}$ articles from recommendations\;
    update $c^{t}$ based on $topk^{t}$\;
    $t \gets t +1$ }
    \label{alg:simulation}
\end{algorithm}

\subsection{Query adaptation for diversification}
\label{sec:diveristy}
One can diversify the recommendation lists by using a fairly similar approach to the method proposed in the previous section. Diversity in news recommendation is essential to burst the filter bubbles around users. In this study, we present a simple approach to increase the number of unique topics/keywords in the articles that are recommended to each user in the simulated rounds. Unlike the fairness-aware hypergraph model (presented in Algorithm~\ref{alg:simulation}), where the coverage weights \(w^{t}_{i}\) are unchanged across users, the diversity weights are defined for the query vector of each user. More specifically, in order to generate news recommendations for user \(u\) we add weights for some randomly sampled~\footnote{The number of samples and the weight are hyperparameters} topics/keywords that have not been recommended to this user in previous rounds. This way we induce diversification to the recommendation lists. This time we do not scale the weights based on the relevance score, as we want to introduce some new and serendipitous topics/keywords to the user.              

\section{Data and experimental setup}
\label{sec4}
In this section we first describe the datasets that we use in the experiments (Section~\ref{sec4_1}). Then we explain how the hyperedges are defined, how the hypergraphs are constructed for each dataset and how simulated rounds are formed (Section~\ref{sec4_2}). Finally in Section~\ref{sec4_3} the performance measures are introduced.

\subsection{Data description}
\label{sec4_1}
We used two news datasets, namely \textit{Roularta}\footnote{Dataset obtained from Roularta Media Group, a Belgian multimedia group.} and \textit{Adressa}~\cite{gulla2017adressa} which are described in Table~\ref{tab:0}, to assess the performance of the proposed hypergraph based RS. These datasets contain metadata information regarding the authors of articles. We consider the authors as the other stakeholders in the system, besides the users. The aim of the proposed model in Section~\ref{sec:query} is to generate fairer recommendations from the authors' perspective. Therefore these datasets are used in Section~\ref{sec5_2} to enhance fairness. The \textit{Roularta} dataset contains title, summary and text information of the articles. To leverage this information, the CNN based deep neural network approach proposed by~\cite{gabriel2019contextual} is used to generate article embeddings for news articles of this dataset. Apart from textual metadata, the \textit{Roularta} dataset also contains IPTC\footnote{International Press Telecommunications Council} tags which are standardized international journalistic tags to facilitate the exchange of news among news agencies. We use the topics (in the \textit{Roularta} dataset) and keywords (in the \textit{Adressa} dataset) as means of diversification in Section~\ref{sec5_3} to evaluate the proposed diversity-aware hypergraph model (Section~\ref{sec:diveristy}). 

\begin{table*}
\centering
  \caption{Datasets descriptions}
    \label{tab:0} 
{  
  \begin{tabular}{lcc}
    \toprule
    & Roularta&Adressa\\
    \midrule
    \# Users &3,899&3,601\\
   \# Items&3,064&710\\
   \# Authors&395&152\\ 
    \# Topics/Keywords&96&855\\  
    \# IPTC tags&422&-\\
   Average \# authors per article&1.05&1.20\\  
    Average \# topics/keywords per article&1.26&5.01\\    
    Average \# IPTC tags per article&5.85&-\\    

   Timespan&60 days& 27 days\\
  \bottomrule
\end{tabular}}
\end{table*}

\begin{figure}
\centering
\begin{subfigure}{.5\textwidth}
  \centering
  \includegraphics[width=\linewidth]{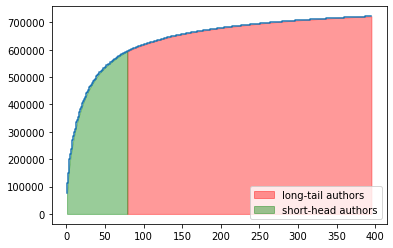}
  \caption{Roularta dataset}
  \label{fig:sub1}
\end{subfigure}%
\begin{subfigure}{.5\textwidth}
  \centering
  \includegraphics[width=\linewidth]{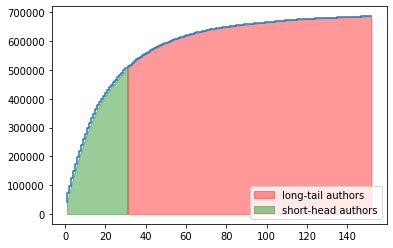}
  \caption{Adressa dataset}
  \label{fig:sub2}
\end{subfigure}
\caption{Authors' cumulative frequencies in interactions}
\label{fig:test}
\end{figure}

Both \textit{Roularta} and \textit{Adressa} datasets have skewed distributions over author popularity in user interactions as illustrated in Figure~\ref{fig:test}. In this figure the X-axis represents author IDs which are sorted descending based on their popularity (number of interactions between users and articles written by these authors) and the Y-axis represents the cumulative popularity values of these authors. According to this figure a small set of authors accounts for a large fraction of the interactions. We call this small set \textit{short-head authors} and the remainings as \textit{long-tail authors}. The \textit{short-head author} group contains the most popular authors, such that their articles account for roughly 20\% of the training data based on the Pareto Principle~\cite{sanders1987pareto}. The other authors belong to the \textit{long-tail author} group. In this study we want to provide fairer recommendations across these two author groups.

\subsection{Experimental setup}
\label{sec4_2}
Based on the available metadata we construct a unified hypergraph for each dataset using the defined hyperedges in Table~\ref{tab:edges}. As is shown in this table, E1 connects each article to the related topics/keywords and the users who have interacted with this article. Similarly, E2 connects each article to the related authors and the users who have interacted with this article. E3 relates each user to the articles that (s)he has interacted with. As we have iptc tags in \textit{Roularta} dataset, each article is connected to the related iptc tags in this dataset via E4. E5 relates each user to his/her \textit{k} nearest neighbor users using the interaction matrix (collaborative information). Finally, E6 connects each article to its \textit{k} nearest neighbor ones based on article embeddings. To calculate the similarity of article embeddings and user interactions we use the cosine similarity measure. For sake of simplicity, we assign an equal weight ($\mathbf{W} = \mathbf{I}$) to all hyperedges and leave hyperedge weight optimization for future work.

\begin{table*}
\centering
  \caption{Hyperedge definitions}
  \label{tab:edges}
  \begin{tabular}{ccccc}
    \toprule
    hyperedge&definition&Roularta&Adressa\\
    \midrule
    \ E1 & article - topics/keywords - users&\checkmark&\checkmark\\
    \ E2 & article - authors - users&\checkmark&\checkmark\\
    \ E3 & user - articles&\checkmark&\checkmark\\
    \ E4 & article - iptc tags&\checkmark&\\
    \ E5 & k similar users&\checkmark&\checkmark\\
    \ E6 & k similar articles&\checkmark&\\
  \bottomrule
\end{tabular}
\end{table*}

To evaluate the performance of the model in the course of time, we form the experiments in simulated rounds. We split each dataset to four simulated rounds based on time. To evaluate the accuracy of the model, for each user~\footnote{We consider users that have more than 10 interactions in the last day.} we hide 10 interactions of that user that occurred in the last day of that round from training in the corresponding round. We used a separate validation set\footnote{This set is formed, similar to forming test set, but using the last day of train set.} to tune the hyperparameter of hypergraph learning \(\vartheta\) as well as the hyperparameters of the compared approaches in Section~\ref{sec5_1}.

\subsection{Performance measures}
\label{sec4_3}
We use four measures to evaluate the performance of the proposed model. To assess the accuracy of the model predictions we use \textit{precision}, which is a standard information retrieval metric. To evaluate recommendation lists w.r.t. fairness two evaluation measures, namely Equity of Attention for group fairness (EAGF)~\cite{mehrotra2018towards}, and Supplier Popularity Deviation (SPD)~\cite{abdollahpouri2020addressing} are used. EAGF is a group fairness measure that can be calculated using Eq.~\ref{eq4}:

\begin{equation}
EAGF = \sum\nolimits_{i=1}^{\abs{g}}\sqrt{\abs{R(i)}}
\label{eq4}
\end{equation}

\noindent where \(g\) is the set of author groups and \(R(i)\) is the set of recommended items that belongs to group \(i\). The higher value in \(EAGF\) means that the RS provides more balanced recommendations (attention) across groups. This measure does not take the real popularity of authors into account. SPD measures the fairness of recommendations based on the popularity of the authors in the dataset using Eq.~\ref{eq5}:     

\begin{equation}
SPD = \frac{\sum\nolimits_{i=1}^{\abs{g}}\abs{\frac{\abs{R(i)}}{\abs{R}}-\frac{\abs{D(i)}}{\abs{D}}}}{\abs{g}}
\label{eq5}
\end{equation}

\noindent where \(R\) is the set of all recommended items, \(D(i)\) is the set of items in the training set that belong to group \(i\) and \(D\) is the set of items in the training set. The lower value of SPD means that the model has better performance in generating calibrated recommendations which are fair and proportional to the real popularity distribution of authors in the training set. To calculate these two fairness measures we need to define author groups. We split authors to long-tail and short-head authors based on their popularity in training data. To evaluate the diversity of content that users have received in their recommendation lists (Section~\ref{sec5_3}), we count the number of unique covered topics/keywords in recommendations till the current simulated round for each user and then report the average value over the users. These four evaluation measures assess the performance based on top 20 recommendations. 

\section{Result and discussion}
\label{sec5}
In this section we report the performance of the proposed methods.  In Section~\ref{sec5_1}, we compare the performance of the (basic) hypergraph RS with some state-of-art RSs w.r.t. accuracy and fairness measures. In Section~\ref{sec5_2}, we assess the fairness-aware hypergraph model that dynamically increases the coverage of under-represented authors and fairness through query vector adaptation. In Section~\ref{sec5_3}, we evaluate the proposed diversity-aware hypergraph based on accuracy and diversity. 

\subsection{Hypergraph ranking for recommendation}
\label{sec5_1}
To demonstrate the merits of hypergraph learning as a fair RS, we compare its performance to some state-of-art RS baselines w.r.t. precision, EAGF and SPD. We apply two versions of the hypergraph, namely with all available data (E1,E2, ..., E6) and without articles metadata (E3, E5). We use matrix factorization for implicit feedback~\cite{hu2008collaborative}, logistic matrix factorization for implicit feedback~\cite{johnson2014logistic}, CNN based content-based filtering~\cite{gabriel2019contextual}, and popularity as well as random based RSs as baselines. As is shown in Table~\ref{tab:b}, even without utilizing the articles metadata, hypergraph (E3,E6) has comparable accuracy compared to the other baselines. Moreover, according to Table~\ref{tab:b} a hypergraph-based RS has better performance in both fairness measures. More precisely, hypergraph-based RSs provide fairer authors' chance of being recommended in recommendations (EAGF) and generate more calibrated recommendations (SPD). As is discussed in Section~\ref{sec:h_theory}, the hypergraph RS is resistant against popularity bias and therefore it gives more calibrated recommendations compared to the other baselines.

\begin{table}[h]
\centering
\captionsetup{justification=centering}
\caption{The performance of hypergraph-based RSs compared to some RS baselines for Roularta dataset}
\label{tab:b}       
\begin{tabular}{lccc}
\hline\noalign{\smallskip}
&Precision $\uparrow$ & EAGF $\uparrow$&SPD $\downarrow$\\
\noalign{\smallskip}\hline\noalign{\smallskip}
Hypergraph (E3,E6)&0.268&249.4&\textbf{0.006}\\
Hypergraph (E1,E2,...,E6)&\textbf{0.305}&247.8&0.028\\
ImplicitMF&0.261&\textbf{249.9}&0.101\\
LogisticMF&0.144&231.2&0.178\\
Content-based&0.112&199.1&0.141\\
Popularity&0.223&221.6&0.206\\
Random&0.065&217.6&0.090\\
\noalign{\smallskip}\hline
\end{tabular}
\end{table}

\subsection{Fairness-aware multi-stakeholder hypergraph-based RS}
\label{sec5_2}
The results of applying the proposed fairness-aware model on the \textit{Roularta} and \textit{Adressa} datasets are shown in Tables~\ref{tab:result-f-roularta} and \ref{tab:result-f-adressa}, respectively. In these tables, we include the performance of three hypergraph based methods, namely \(hypergraph\), \(hypergraph\_coverage\) and \(hypergraph\_fair\) in simulated rounds. \(hypergraph\) is the hypergraph model without any adaptation for coverage and fairness. \(hypergraph\_fair\) is the proposed model illustrated in Algorithm~\ref{alg:simulation}. We include \(hypergraph\_coverage\) to show that considering only the coverage without taking into account the relevance of the authors in recommendations will have a considerable negative effect on the accuracy of the model. In this model we do not consider the authors' relevance scores (\(r^{t}\)) in adapting the authors' weights in the query vectors. We also include a simulation-based multi-stakeholder coverage re-ranking baseline~\cite{burke2016towards} in which the initial recommendation lists generated by the \textit{hypergraph} model will be re-ranked based on the stakeholders' (authors) coverage.       

The performance of the fairness-aware model based on the \textit{Roularta} dataset is shown in Table~\ref{tab:result-f-roularta}. In this table, each of the baselines has relatively lower accuracy compared to \(hypergraph\) as they try to increase at least one of the fairness measures at the cost of reduced accuracy. The \(coverage\_reranking \) approach can improve the EAGF measure considerably but it makes the recommendations less calibrated and therefore deteriorate the SPD measure. The proposed model (\(hypergraph\_fair\)) has the best performance in SPD and the lowest drop in precision compared to the other baselines. It can also improve EAGF compared to \(hypergraph\). We include the \(hypergraph\_coverage\) performance to show that improving fairness without considering relevance has a considerable negative effect on accuracy and SPD.

\begin{table*}[h]
\centering
  \caption{Results of fair news recommendation for Roularta dataset}
  \label{tab:result-f-roularta}
  \begin{tabular}{llcccc}
    \toprule
    Measure&Option&Sim1&Sim2&Sim3&Sim4\\
    \midrule
    
    \multirow{2}{*}{Precision $\uparrow$}&hypergraph & 0.223&0.105&0.123&0.143\\
    &hypergraph\_fair & 0.223&\textbf{0.09}5&\textbf{0.085}&\textbf{0.124}\\
&hypergraph\_coverage & 0.223&0.014&0.028&0.033\\
&coverage\_reranking & 0.223&0.081&0.032&0.067\\
    \midrule
    \multirow{2}{*}{EAGF $\uparrow$}&hypergraph &272.4&378.6&464.0&533.7\\
    &hypergraph\_fair &272.4&388.1&484.4&558.8\\
&hypergraph\_coverage &272.4&439.1&\textbf{547.0}&\textbf{623.5}\\

&coverage\_reranking &272.4&\textbf{440.7}&541.5&622.8\\
    \midrule
    \multirow{2}{*}{SPD $\downarrow$}&hypergraph & 0.060&0.109&0.142&0.157\\
    &hypergraph\_fair & 0.060&\textbf{0.036}&\textbf{0.027}&\textbf{0.013}\\
&hypergraph\_coverage & 0.060&0.424&0.526&0.615\\

&coverage\_reranking &0.060&0.148&0.378&0.520\\
  \bottomrule
\end{tabular}
\end{table*}

The result of applying the proposed fairness-aware model on the \textit{Adressa} dataset is shown in Table~\ref{tab:result-f-adressa}. Similar to the \textit{Roularta} dataset, in this table \(hypergraph\_fair\) has again relatively better performance in SPD and also EAGF with a lower drop in accuracy. \(coverage\_reranking \) has the best performance in EAGF but with a very high cost of reduced accuracy and worse SPD. 

\begin{table*}[h]
\centering
  \caption{Results of fair news recommendation for Adressa dataset}
  \label{tab:result-f-adressa}
  \begin{tabular}{llcccc}
    \toprule
    Measure&Option&Sim1&Sim2&Sim3&Sim4\\
    \midrule
    
    \multirow{2}{*}{Precision $\uparrow$}&hypergraph&0.246&0.174&0.138&0.119 \\
    &hypergraph\_fair &0.246&\textbf{0.173}&0.126&\textbf{0.109}\\
&hypergraph\_coverage&0.246&\textbf{0.173}&\textbf{0.132}&0.067\\

&coverage\_reranking &0.246&0.171&0.074&0.077 \\
    \midrule
    \multirow{2}{*}{EAGF $\uparrow$}&hypergraph & 416.5&729.6&829.1&695.6\\
    &hypergraph\_fair & 416.5&731.0&834.0&700.8\\
&hypergraph\_coverage & 416.5&728.7&829.5&687.9\\
&coverage\_reranking &416.5&\textbf{733.8}&\textbf{877.0}&\textbf{845.4}\\
    \midrule
    \multirow{2}{*}{SPD $\downarrow$}&hypergraph&0.012&0.024&0.127&0.268 \\
    &hypergraph\_fair&0.012&\textbf{0.004}&\textbf{0.090}&0.257 \\
&hypergraph\_coverage&0.012&0.020&0.115&\textbf{0.251}\\
&coverage\_reranking&0.012&0.029&0.243&0.298\\
  \bottomrule
\end{tabular}
\end{table*}

The proposed approach (\(hypergraph\_fair\)) has relatively lower drop in accuracy compared to the other baselines in \textit{Roularta} and \textit{Adressa} datasets. This approach has the best performance in SPD as it is designed to have calibrated recommendations. This indicates that we lose a little bit of accuracy in order to have fairer and more calibrated recommendations from the viewpoint of other stakeholders. A fairer recommendation does not mean that all the authors have exactly the same exposure. Authors that write many articles have still a higher exposure to ones that are not very active. However, our proposed approach succeeds in mitigating the popularity bias and increases the coverage of under-represented authors.

\subsection{Diversity-aware hypergraph RS}
\label{sec5_3}
The results of applying the proposed diversity-aware model on the \textit{Roularta} and \textit{Adressa} datasets are shown in Tables~\ref{tab:result3} and~\ref{tab:result4}, respectively. In these tables we include the performance of \(hypergraph\), \(hypergraph\_diversified\) and \(diversity\_reranking\) in four simulated rounds. \(hypergraph\) is the hypergraph model without any adaptation for diversity, \(hypergraph\_diversified\) is the proposed model explained in Section~\ref{sec:diveristy} and \(diversity\_reranking\) is the adapted version of~\cite{burke2016towards} to increase topics/keywords coverage. As is shown in these tables, \(hypergraph\_diversified\) can expand the set of recommended topics/keywords to the readers with only a small accuracy drop in the course of time. In the \textit{Roularta} dataset on average 17\% improvement in diversity came with 6\% drop in precision while in the \textit{Adressa} dataset 10\% percent expansion in the set of recommended keywords accounts for 4.5\% lower \textit{precision}. For both datasets, \(diversity\_reranking\) considerably improves diversity however, it drastically deteriorates precision. Since the very essential role of an RS is to provide relevant recommendations, this high drop in \textit{precision} certainly jeopardizes users' experience.

\begin{table*}[h]
\centering
  \caption{Results of news recommendation for Roularta dataset (diversification)}
  \label{tab:result3}
  \begin{tabular}{llcccc}
    \toprule
    Measure&Option&Sim1&Sim2&Sim3&Sim4\\
    \midrule
    
    \multirow{2}{*}{Precision $\uparrow$}&hypergraph&0.223&0.105&0.123&0.143 \\
&hypergraph\_diversified &0.223&0.098& 0.118&0.132\\
&diversity\_reranking &0.223&0.086&0.063&0.070\\
    \midrule
        \multirow{2}{*}{covered\_topics $\uparrow$}&hypergraph & 7.7&8.2&8.9&9.8\\
&hypergraph\_diversified & 7.7&9.9&9.5&12.0\\
&diversity\_reranking&7.7&18.9&23.6&24.7\\
    \midrule
\end{tabular}
\end{table*}

\begin{table*}[h]
\centering
  \caption{Results of news recommendation for Adressa
  dataset (diversification)}
  \label{tab:result4}
  \begin{tabular}{llcccc}
    \toprule
    Measure&Option&Sim1&Sim2&Sim3&Sim4\\
    \midrule
    
    \multirow{2}{*}{Precision $\uparrow$}&hypergraph&0.246&0.174&0.138&0.119 \\
&hypergraph\_diversified &0.246&0.165&0.128&0.117\\
&diversity\_reranking &0.246&0.131&0.105&0.082\\
    \midrule
        \multirow{2}{*}{covered\_keywords $\uparrow$}&hypergraph & 34.7&30.5&35.5&41.9\\
&hypergraph\_diversified &34.7&32.6& 42.7&42.7\\
&diversity\_reranking&34.7&46.4&42.9&44.5\\
  \bottomrule
\end{tabular}
\end{table*}

\section{Conclusion}
\label{sec6}
The main contribution of this paper is addressing fairness issue in a multi-stakeholder news recommendation setting using hypergraph learning. In this setting, we enhanced the coverage of under-represented stakeholders (in our case, authors) in the recommendation lists by dynamically adapting their weights in the query vectors based on their popularity in the users' interactions and their coverage in the previous rounds of recommendations. Although this will come with the cost of reduced accuracy it has a big impact on the coverage of under-represented authors and generates fairer recommendations towards non-popular authors. We also showed that, with a relatively similar approach, we are able to diversify the recommendations in the simulated rounds by expanding the set of topics/keywords that are recommended to the users.

An interesting topic for future work would be to extend this study by including time decays for interaction weights in the hypergraph. In news aggregator websites the recent articles are usually more relevant compared to the very old ones. This can help a hypergraph-based RS to provide more accurate recommendations. Moreover, the proposed fairness-aware hypergraph can be applied to other types of recommender systems, such as movie ones. In movie recommendations, a hypergraph-based RS can be used to generate fair recommendations from the perspectives of directors, producers, as well as cast members.

\section{Acknowledgements}
This work was executed within the imec.icon project NewsButler, a research project bringing together academic researchers (KU Leuven, VUB) and industry partners (Roularta Media Group, Bothrs, ML6). The NewsButler project is co-financed by imec and receives project support from Flanders Innovation \& Entrepreneurship (project nr. HBC.2017.0628). The authors also acknowledge support from the Flemish Government (AI Research Program).

\bibliography{mybibfile}

\end{document}